**Title:** Clustering indications before the $M_w$7.0 2020 Samos, Greece, main shock as revealed in an equivalent dimensions space

**Authors:**

Stanislaw Lasocki, Institute of Geophysics Polish Academy of Sciences

Vasileios G. Karakostas, Geophysics Department, Aristotle University of Thessaloniki

Eleftheria E. Papadimitriou, Geophysics Department, Aristotle University of Thessaloniki

**Corresponding author:** Stanislaw Lasocki, Institute of Geophysics Polish Academy of Sciences, 64 Księcia Janusza Str., 01-452 Warszawa, Poland lasocki@igf.edu.pl

**Conflict of interest statement:** The authors declare that there are no conflicts of interest recorded.

# Abstract

The transformation to equivalent dimensions that offers a novel approach for investigating earthquake clustering was engaged to analyze the preparatory phase of the 2020 Samos, Greece, $M_w$7.0 main shock. The analysis considered earthquakes that occurred between 2006 and October 2020, covering an area extended three times the length of the main rupture. Each earthquake was parameterized by its magnitude, the interevent time (interval since the previous earthquake), and the interevent spatial distance (distance between the epicenters of consecutive earthquakes). Transforming these parameters into equivalent dimensions allowed them to be directly compared. The degree of clustering was quantified using the average distance between earthquakes in this transformed parameter space, calculated within consecutive 100–events data windows. Results revealed a distinct pattern: the average distance was increasing steadily during the twelve-year period before the main shock. These temporal changes in the average distance were driven by a systematic evolution of earthquake clustering in the used parameter space. Beginning from a two-cluster system, when the distance was minimal, the clustering development continued along two branches and ended before the main shock with the formation of five earthquake clusters of different characteristics.

# Introduction

Earthquakes, likely the most catastrophic expression of Earth's dynamics have often very high death toll, leave tens to hundreds of thousands homeless, and set countries' economies back decades. Any new solutions that could contribute to reducing losses caused by earthquakes are of great significance to science and society. However, the seismic process is multidimensional and highly nonlinear. Due to that, a reliable prediction of the time, location and magnitude of earthquakes remains impossible. The earthquake seismology aim – "earthquake prediction" was reformulated in 2007 during the 5[th] International Workshop on Statistical Seismology to "earthquake forecasting," understood as providing estimates of the time – variant probabilities of earthquakes occurring within some region (Jordan et al., 2007).

Many authors indicate that studies of earthquake clustering that reflect stochasticity of seismic processes may improve earthquake predictability (e.g., Zaliapin and BenZion, 2021; Kato and Ben-Zion, 2021, Misrah et al., 2024, and the references therein). Lasocki et al. (2024) investigated clustering behavior of the seismic process preceding the 2017 $M_w$8.2 Thuantepec, Mexico main shock. Instead of usually exploited occurrence time and focal coordinates, they traced clustering evolution in the space of waiting time, epicentral offset and magnitude. The earthquake waiting time is the time since the previous earthquake after which this earthquake occurred and the earthquake epicentral offset is the distance of the epicenter of this earthquake from the epicenter of the previous earthquake. The transformation of these three parameters to equivalent dimensions (ED-s, Lasocki, 2014) ensured their mutual comparability and the Euclidean metric in the ED space. The average Euclidean distance between earthquakes in the space of these three transformed parameters was used to determine the degree of earthquake clustering. The time–changes of clustering were expressed by the time-changes of this average distance calculated in sliding data windows. The results of this study were very interesting. Some ten years before the main shock, the events grouped in one cluster of minimum average distance between earthquakes. Next, the distance was increasing until two years before the main shock, and then, it declined until the main shock. This precursory down-up–down signal was highly significant statistically.

The Tehuantepec earthquake was large and, like M8+ earthquakes usually do, ruptured the entire lithosphere. Studies of other large earthquake cases are on the way and the preliminary results are also very promising. To see whether the approach used in Lasocki, et al. (2024) can be fruitful also for smaller earthquakes, here we applied it for the main shock originating in different, much complex tectonic regime and of different magnitude class. Our choice affiliated the 2020 Samos main shock, with $M_w$7.0, which occurred in the back arc Aegean area. Intraplate earthquake forecasting is more challenging than for their interplate counterparts, given the much longer recurrence times along with uncertainties in estimating the strain accumulation on the causative faults. The 2020 main shock is among the strongest earthquakes in the Aegean area that are recorder during the instrumental era, and this makes the identification of any premonitory pattern quite challenging. Our results showed that the main shock occurrence was preceded by a statistically significant minimum of the average distance between earthquakes at about 12 years before the main shock and then by a systematic increase of this distance until the main shock. Like in the mentioned Tehuantepec earthquake case, there was no premonitory signal when clustering was studied in the spaces of original earthquake parameters.

## Materials and Methods

### Tectonic Background of the Aegean Area and the 2020 Samos main shock

The 30 October 2020 $M_w$7.0 Samos main shock took place offshore north of Samos Island in eastern Aegean area, previously struck in 1904 with a comparable magnitude earthquake offshore the southern coastline (red star Figure 1). Several historical $M \geq 6.0$ earthquakes are reported for the area around Samos Island and the nearby western Turkish coastline known from historical records since 496 BC (Papazachos and Papazachou, 2003). The map of Figure 1 shows their epicentral distribution (stars with the occurrence year next to each symbol) in both affected areas. This distribution evinces that both areas were very frequently struck by strong earthquakes even though the historical catalog is not complete since the beginning of the macroseismic archiving and evaluation.

Near Samos Island in particular, the historical catalog includes the 200 BC M6.3 (max. intensity in MKS scale VIII in Samos), the 47 AD M6.9 (VIII in Samos), the 1751 M6.9 (VIII in Samos), the 1831 M6.0 (VII in Samos), the 1846 M6.0 (VII in Samos), the 1865 M6.0 (VII in Samos), the 1868 M6.0 (VII in Samos), the 1873 M6.9 (VIII in Samos), and the 1893 M6.9 (VII in Samos). If we assume that the historical record is almost complete for $M \geq 6.0$ earthquakes in the second half of the 19[th] century, we may speculate that the northern fault segments of Samos Island could be associated with strong main shocks in 1873 (seriously harming the eastern part of Samos Island and strongly felt in the city of Izmir, Turkey) and in 1877 (when the most severe damage was reported for the north part of the island near the town of Samos). Soon afterwards, in 1890 and in 1893 Samos, the nearby Turkish coastal area underwent damage. It is remarkable to observe how tightly clustered the strong earthquakes appear, with the consecutive failure of adjacent fault segments in just ten years.

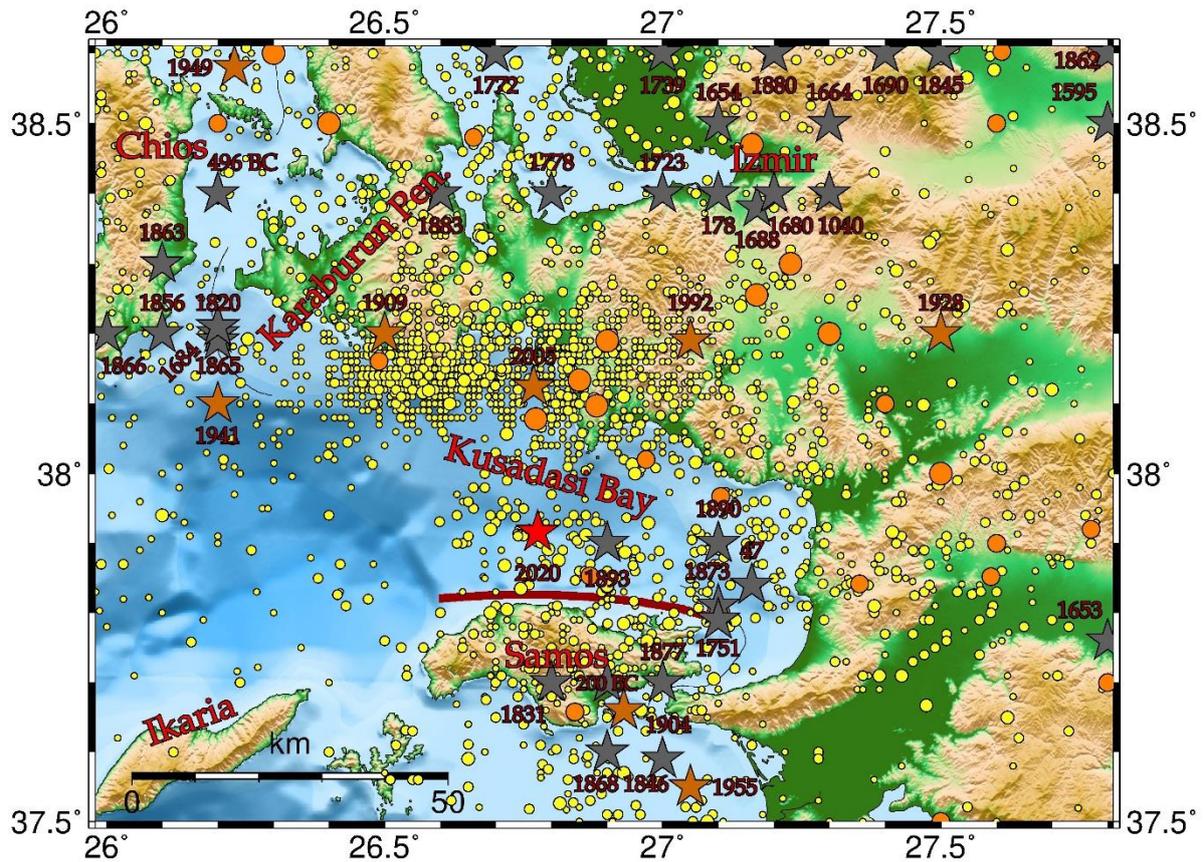

*Figure 1. The area of eastern Aegean Sea, the epicentral distribution of historical (496BC–1893) and instrumental (1900–present) large (M≥6.0) earthquakes are depicted by the gray and orange stars, respectively, and the epicenter of the 2020 mainshock by a red star. Strong earthquakes (5.0≤M<6.0) epicenters are shown by the orange circles and moderate seismicity (3.0≤M<5.0) by smaller yellow circles (data from http://geophysics.geo.auth.gr/ss/; doi:10.7914/SN/HT).*

The study area belongs to the eastern Aegean extensional regime that incorporates the islands and offshore areas along with western Anatolia. Normal faulting is the dominant style of faulting, where the relatively high N–S stretching deformation localized in almost E–W striking faults, along with considerable activity on secondary oblique fault segments, capable to produce considerable seismicity rates and frequent strong earthquakes (Papazachos et al., 1998; Kassaras et al., 2000; among others). Normal faulting due to the intense stretching of the over thrusting Aegean microplate caused by the slab's roll back, namely the subduction of oceanic crust of the eastern Mediterranean (Papazachos and Comninakis, 1971; McKenzie, 1972). Geodetic and seismological data suggest slip rates up to 5 mm yr$^{-1}$ on major regional faults (McClusky et al., 2000).

The part of the eastern Aegean area where the 2020 Samos main shock took place belongs to a basin located north of Samos Island. The global centroid moment tensor solution (GCMT; www.globalcmt.org) gave a $M_o$=4.01*10$^{19}$ N·m ($M_w$7.0) with a centroid depth of 12 km and duration of 15.4 s. It is an almost pure normal faulting with fault plane parameters given by strike, $\varphi_1=270°$, dip, $\delta_1=37°$, and rake, $\lambda_1=-95°$, and $\varphi_2=96°$, $\delta_2=53°$, $\lambda_2=-86°$. The aftershock sequence of the 2020 Samos main shock was exhaustively investigated with a highly accurate aftershock catalog, fault plane solutions and Coulomb stress changes calculation (Karakostas et al., 2021). The vast majority of the preliminary aftershock locations is concentrated in the offshore area north of Samos Island. Their epicentral alignment agrees well with the normal faulting type of the almost E–W striking main rupture, the distance lengthwise is, nevertheless, rather larger than what is expected for a $M_w7.0$ main shock, which is estimated equal to 42 km from empirical relations for normal faults by both Wells and Coppersmith (1994) and Papazachos et al. (2004). The highly accurate epicentral locations evidence the activation of multiple secondary fault segments comprising the local fault network (Karakostas et al., 2021).

## Data

We have taken our data set from the regional earthquake catalog compiled in the Geophysics Department of the Aristotle University of Thessaloniki (http://geophysics.geo.auth.gr/ss/catalogs_en.html), based on the recordings of the Hellenic Unified Seismological Network (HUSN). We considered a rectangular area with length three times the fault length of the 2020 Samos main shock (red star in Figure 2), namely one fault length either side beyond the tips of the main rupture. This region is regarded as the area of direct mechanical coupling because the lobes of positive Coulomb stress changes extend approximately one fault length beyond the edges of the main rupture. The width of the rectangle was selected to exclude seismicity associated with other fault populations.

It comprises 917 crustal (h≤40 km) earthquakes with magnitudes in the range $3.0≥M_w≥7.0$ that occurred from 18 April 1980 up to the end of October 2020. The magnitudes in the catalog are expressed in the moment magnitude scale ($M_w$) or as equivalent $M_w$ values, derived using scaling relations proposed by Papazachos et al. (1997b). We investigated the catalog completeness in varying time intervals and finally we concluded with $M_c=3.0$ in 2006–2020, which resulted in a data set comprising 372 earthquakes.

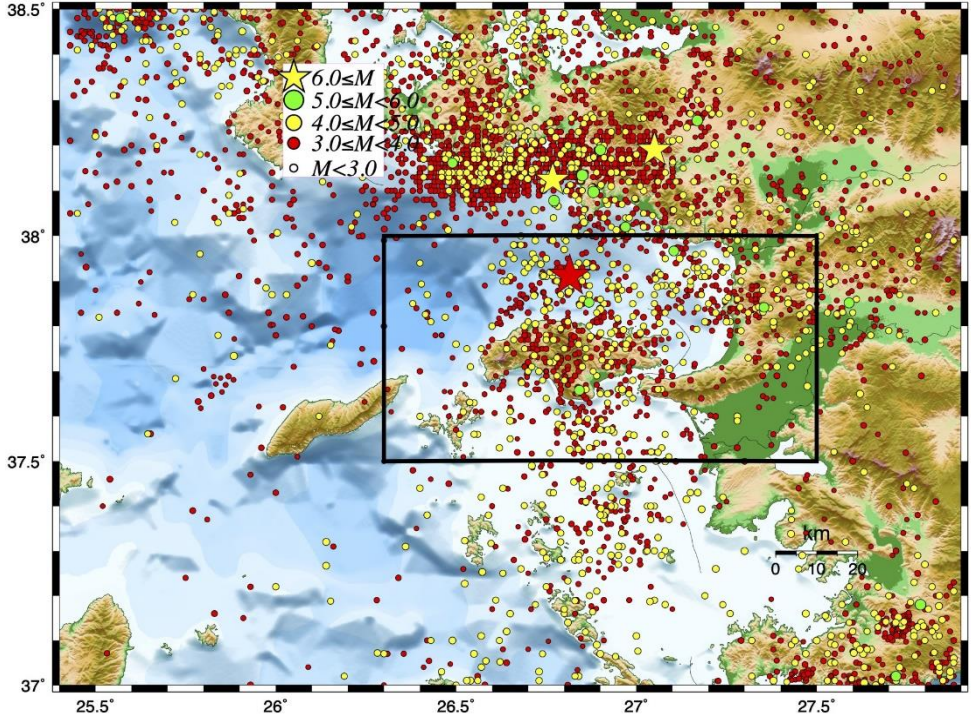

*Figure 2. Instrumental seismicity between 1980 and October 2020. Symbols are according to earthquake magnitude as indicated in the inset. The red star depicts the 2020 Samos main shock epicenter.*

## Method

The data was analyzed using the same method as the one in the Lasocki et al., 2024 paper. Therefore, we provide here only a brief description of this method, and all the details may be found in the cited paper.

The data processing consisted of the following six steps:

Step 1. Estimating the magnitude completeness level and extracting the complete part from the initial dataset. The completeness level was calculated using the modified goodness-of-fit test (Leptokaropoulos et al., 2013).

Step 2. Reparameterization of seismic events from the extracted seismic series preceding the main shock. For an event $k$ its new parameters were the waiting time, $dt_k = t_k - t_{k-1}$, where $t_\bullet$ was the event occurrence time, the epicentral offset, $dr_k$ i.e., the orthodromic epicentral distance between the events $k$ and $k$-1, and the event magnitude, $M_k$.

Step 3. Transformation of $\{dt_k, dr_k, M_k\}$ to equivalent dimensions $\{DT_k, DR_k, MC_k\}$. The transformation to equivalent dimension (Lasocki, 2014; 2021) replaces parameter values with their cumulative distribution function (CDF) values. Unknown CDFs of the parameters under study are estimated from data by the non-parametric kernel estimation method (Silverman, 1986). All parameters transformed to EDs are uniformly distributed in [0, 1], hence comparable, and the space of ED–transformed parameters has the Euclidean metric.

Step 4. Calculating the average distance between earthquakes parameterized by *DT*, *DR*, *MC* in a data window of constant size sliding over the event series preceding the main shock. For an *n*-earthquake window the average distance is:

$$d_c = \frac{2}{(n-1)n}\sum_{k=1}^{n-1}\sum_{j=k+1}^{n}\sqrt{(DT_k - DT_j)^2 + (DR_k - DR_j)^2 + (MC_k - MC_j)^2}. \tag{1}$$

$d_c$ was linked to the occurrence time of the last earthquake in its window.

Step 5. A comparative study of earthquake clustering in spaces of original earthquake parameters. First, the earthquake occurrence time, epicentral coordinates, and magnitude were transformed to equivalent dimensions: *TC*, *LAT*, *LON*, *MC*, as in step 3. Then, the average distance between earthquakes (as in step 4) was calculated separately in the {*TC*, *LAT*, *LON*} and {*TC*, *LAT*, *LON*, *MC*} spaces.

Step 6. Insight into the evolution of earthquake clustering in the {*DT*, *DR*, *MC*} space, linked to the time-variation of $d_c$. The *k*-means clustering method was used to divide earthquakes from selected data windows into a number of clusters.

Step 6.1 Assessing the optimal number of earthquake clusters in the selected data windows. The gap optimality criterion with the principal component analysis (PCA) reference data-generation method was used (Tibshirani et al., 2001).

Step 6.2 Dividing earthquakes from the selected data windows of into the optimal number of clusters, and calculating clusters' parameters: the centroid coordinates, number of cluster members, distances between centroids, and mean squared distance of cluster members from the centroids.

## Results and Discussion

The seismic data were not homogeneous concerning the magnitude completeness level. For all data from 1980-2020 the completeness level was high, equal to 4.1 whereas for the data from 2006-2020 it equals 3.0. Following, the complete part of the whole data sample (1980-2020) comprised 233 events, and the complete part of the data from 2006-2020 was better populated amounting to 372 events. For this reason, we used the series of earthquakes from 2006 to 2020 as data for the analysis.

The time changes of the average distance between earthquakes in the {*DT*, *DR*, *MC*} space, $d_c(t \mid DT, DR, MC)$ while the main shock was approaching are shown in Figure 3. This result was obtained using the data window of 100 events being consecutively shifted of one event. The presented $d_c$ – values are linked to the times of the last events in windows, respectively. The time of the last value of $d_c$ is the occurrence time of the last event before the main shock. The solid horizontal line represents the mean value of $d_c$ when one hundred [*DT*, *DR*, *MC*] points in a window have the uniform distribution in a 3D [0.1] cube, and the horizontal dashed line are the limits of the 90-percent confidence interval $d_c$ for a single draw of one hundred such points in the window. These mean value and confidence interval limits are 0.662, 0.632, and 0.691, respectively (see: Lasocki et al., 2024 for details)

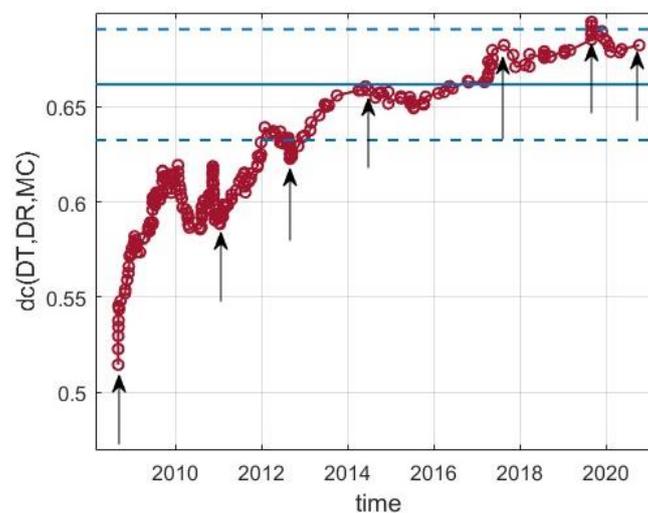

*Figure 3. Time series of the average distance between a hundred consecutive earthquakes in the {equivalent waiting time, DT; equivalent epicentral offset, DR; equivalent magnitude, MC} space. Solid and dashed horizontal lines are the mean value, <$d_c$>, and limits of 90–percent confidence interval of $d_c$ for a single 100-element draw of uniformly distributed points. Arrows indicate the $d_c$ values with which the data windows used to study the evolution of earthquake clustering are associated.*

As can be seen, $d_c(t \mid DT, DR, MC)$ begins from a low value at 1/09/2008 (the occurrence date of the last event in the first window), much lower than statistical possibilities for $d_c$ scatter when the [*DT*, *DR*, *MC*] points are uniformly distributed. Due to the time limit of the analyzed data that started from 2006 it is unclear whether this value is an actual minimum or the minimum is sometime before the first data point of $d_c$. Nevertheless, this $d_c$ – value signifies that in that window the earthquakes – the [*DT*, *DR*, *MC*] points are strongly clustered.

With time, $d_c$ distinctly systematically increases, getting the value 0.695 in the window number 259 linked to 30/09/2019 20:15:42. This maximum value is above the upper limit of 90–percent confidence interval for $d_c$ for earthquakes uniformly distributed in the window. In the last period before the main shock $d_c$ drops down to 0.682.

The increasing trend of $d_c$ in the period before the main shock, with a slight decrease or plateau of DC in the last period just before the main shock, is similar to the course of $d_c$ before the Tehuantepec earthquake, the one shown in Figure 4. Only the drop–down of $d_c$ before the main shock, after $d_c$ reaches the maximum, is less pronounced in the Samos earthquake case. The 2017 M8.2 Tehuantepec and 2020 M7.0 Samos earthquakes strongly differed in magnitude and complexity of seismotectonic settings. In spite of that, in both cases, $d_c$ exhibited similar clearly identifiable signals before the main shocks.

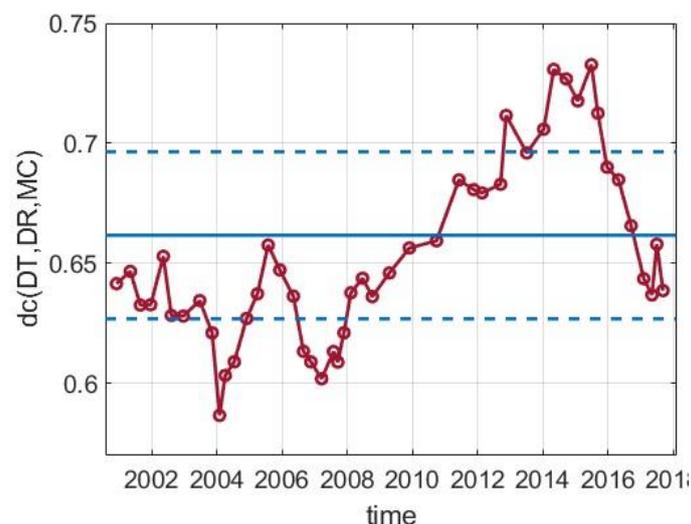

*Figure 4 Time series of the average distance between a hundred consecutive earthquakes in the {equivalent waiting time, DT; equivalent epicentral offset, DR; equivalent magnitude, MC} space before the 2017 M8.8 Tehuantepec main shock. From Lasocki et al., 2024, modified.*

If calculated in the spaces of original earthquake parameters transformed to equivalent dimensions, {*TC*, *LAT*, *LON*}, {*TC*, *LAT*, *LON*, *MC*}, where *TC* stands for earthquake occurrence time transformed to ED, no precursory signal of $d_c$ is present, what is illustrated in Figure 5. All $d_c(t \mid TC, LAT, LON)$ values are much below the lower limit of 90–percent confidence interval of $d_c$ for one hundred [*TC*, *LAT*, *LON*] points uniformly distributed in the [0, 1] 3D cube. From one million draws of one hundred points uniformly distributed in the [0, 1] 4D cube, we established that the mean value of $d_c$ and its 90-percent confidence interval for a single such draw are 0.778 and (0.748, 0.807), respectively. $d_c(t \mid TC, LAT, LON, MC)$ values are much below the lower limit of this 90–percent confidence interval. The earthquakes were clustered in the {*TC*, *LAT*, *LON*} and {*TC*, *LAT*, *LON*, *MC*} spaces throughout the analyzed period, but in both spaces the degree of clustering did not vary significantly when approaching the main shock occurrence time. The same situation of no precursory clustering changes in the original earthquake parameters spaces was met in the Tehuantepec earthquake case (Lasocki et al., 2024). In both Tehuantepec and Samos cases, significant $d_c$ variations before the main shocks has appeared only when the preceding seismicity has been parameterized by the waiting time, $DT$, epicentral offset, $DR$, and magnitude, $MC$. When the original earthquake parameters have been retained (occurrence time, latitude, longitude, magnitude), $d_c$ has not provided any precursory signal.

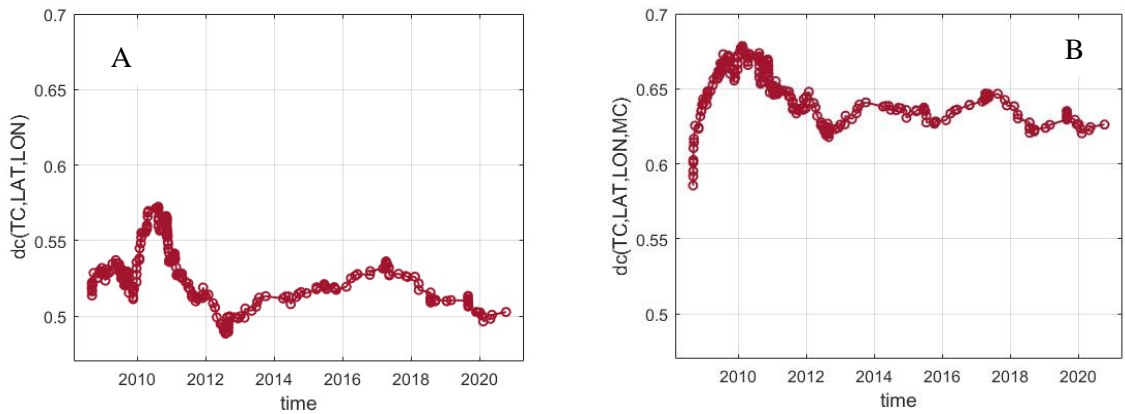

*Figure 5. Time series of the average distance between a hundred consecutive earthquakes in A – the {equivalent occurrence time, TC; equivalent epicenter latitude, LAT; equivalent epicenter longitude, LON} space, and B – the {equivalent occurrence time, TC; equivalent epicenter latitude, LAT; equivalent epicenter longitude, LON, equivalent magnitude MC} space. The vertical scale matches the vertical scale in Figure3, enabling comparisons of $d_c$ variations in **both figures**.*

To get an insight into the evolution of clustering in the {$DT$, $DR$, $MC$} space linked to the variation of $d_c$ we analyzed the optimal number of clusters in each of seven windows indicated by arrows in Figure 3. Then, we divided the earthquakes, represented by the [$DT$, $DR$, $MC$] points, in these windows into the optimal number of clusters for the particular windows. Table 1 provides the results of this analysis.

*Table 1. Results of the division of earthquakes in windows into the optimal number of clusters.*

| No. | Window time point, $t$ | $d_c(t)$ | Optimal number of clusters | Cluster No | Centroid coordinates | | | Number of cluster's members |
|---|---|---|---|---|---|---|---|---|
| | | | | | $DT$ | $DR$ | $MC$ | |
| 1 | 1/09/2008 | 0.514 | 2 | I | 0.31 | 0.32 | 0.77 | 37 |
| | | | | II | 0.61 | 0.71 | 0.73 | 63 |
| 2 | 11/01/2011 | 0.588 | 2 | I | 0.60 | 0.76 | 0.34 | 36 |
| | | | | II | 0.26 | 0.26 | 0.37 | 64 |
| 3 | 28/08/2012 | 0.625 | 3 | I | 0.23 | 0.22 | 0.24 | 44 |
| | | | | II | 0.72 | 0.69 | 0.36 | 32 |
| | | | | III | 0.35 | 0.26 | 0.72 | 24 |

| | | | | | | | |
|---|---|---|---|---|---|---|---|
| 4 | 2/06/2014 | 0.661 | 2 | I | 0.29 | 0.22 | 0.41 | 59 |
| | | | | II | 0.77 | 0.71 | 0.44 | 41 |
| 5 | 18/08/2017 | 0.682 | 4 | I | 0.82 | 0.62 | 0.22 | 31 |
| | | | | II | 0.16 | 0.17 | 0.30 | 25 |
| | | | | III | 0.81 | 0.80 | 0.67 | 23 |
| | | | | IV | 0.54 | 0.12 | 0.64 | 21 |
| 6 | 30/08/2019 | 0.695 | 5 | I | 0.13 | 0.15 | 0.31 | 22 |
| | | | | II | 0.87 | 0.71 | 0.25 | 25 |
| | | | | III | 0.29 | 0.09 | 0.75 | 15 |
| | | | | IV | 0.82 | 0.73 | 0.71 | 24 |
| | | | | V | 0.74 | 0.21 | 0.27 | 14 |
| 7 | 9/10/2020 | 0.682 | 5 | I | 0.86 | 0.74 | 0.23 | 22 |
| | | | | II | 0.15 | 0.17 | 0.29 | 24 |
| | | | | III | 0.85 | 0.69 | 0.69 | 26 |
| | | | | IV | 0.74 | 0.21 | 0.27 | 14 |
| | | | | V | 0.28 | 0.10 | 0.71 | 14 |

The observed part of the preparatory process begins from two clusters in window 1. However, compared with other two cluster divisions in windows 2 and 4, the two clusters of window 1 are closer to each other and denser. The distances between centroids and the mean squared distance of clusters members from their centroids in windows 1, 2, and 4 are 0.50, 0.60, and 0.68 and 0.10, 0.12, and 0.14, respectively. The division into two clusters in window 1 probably follows a single cluster of the earlier (unobserved) data group.

The beginning with two clusters in window 1 ends with five clusters in window 7 – the last data window before the main shock. Based on locations of cluster centroids in windows (Table 1), we identify five paths of an evolutionary process of earthquake clustering as schematically presented in Figure 6 and Table 2. The values of centroid coordinates are shown in Figure 7.

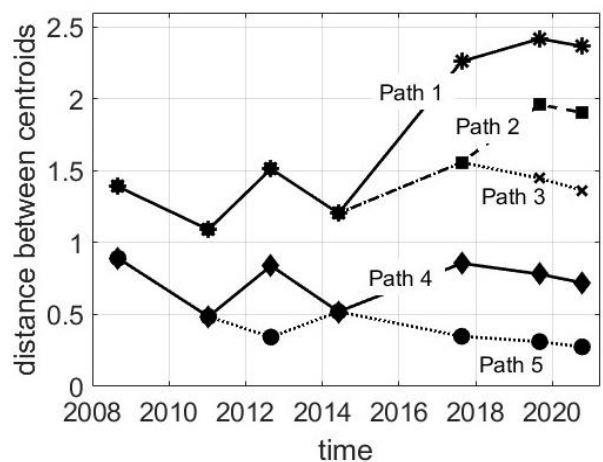

*Figure 6. Schematically picture of the earthquake clustering process in the {waiting time, DT; epicentral offset, DR; magnitude, MC} equivalent dimensions space preceding the 2020 Mw 7.0 Samos earthquake. y-value of the first lower point (diamond) is the distance the location of the cluster I centroid from the (0, 0, 0) point. y-value of each of the next lowest points in the*

*following windows is equal to the y-value of the preceding lowest point plus or minus the distance between the centroid linked to the current point and the centroid linked to the preceding point. Other points in every window are in the order of increasing distances of their centroids from the centroid of the lowest point. The y-value of every such point is equal to the y-value of the directly lower point and the distance between the centroids linked to these two points.*

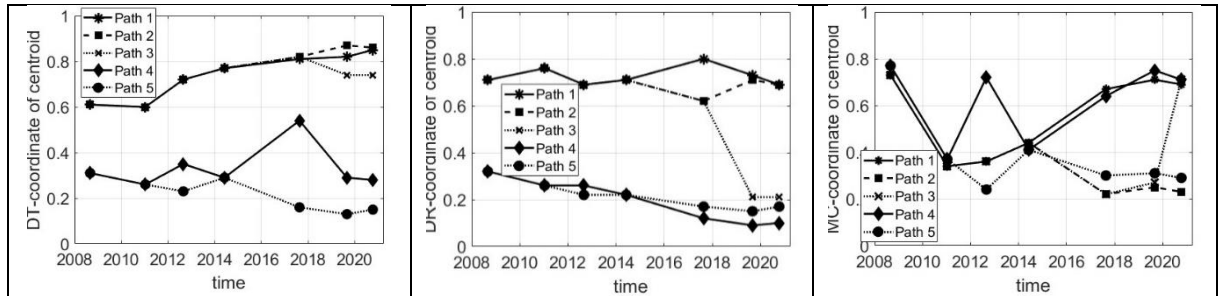

*Figure 7. Coordinates of cluster centroids in consecutive windows.*

*Table 2. Identified paths of the earthquake clustering process in the {waiting time, DT; epicentral offset, DR; magnitude, MC} equivalent dimensions space preceding the 2020 Mw 7.0 Samos earthquake*

| Window No: | 1 | 2 | 3 | 4 | 5 | 6 | 7 |
|---|---|---|---|---|---|---|---|
| Path | Cluster No in window | | | | | | |
| 1 | II | I | II | II | III | IV | III |
| 2 | II | I | II | II | I | II | I |
| 3 | II | I | II | II | I | V | IV |
| 4 | I | II | III | I | IV | III | V |
| 5 | I | II | I | I | II | I | II |

The division into two clusters in window 1 continues as two different branches of clusters throughout the whole period. The cluster in window 4 of the first upper branch in Figure xx4 splits into two clusters in window 5, and in window 6 a three-cluster structure is formed that continues in window 7. The lower branch cluster splits in window 2 splits into two clusters in window 3, that unite temporarily in window 4. Then again, two clusters in window 5 are built and this division is retained up to window 7.

The earthquakes in window 1 are divided into two clusters whose *DT* and *DR*-coordinates of centroids are larger and small, respectively. The earthquakes from those clusters had longer waiting times and long epicentral offsets, and shorter waiting times and shorter epicentral offsets. The magnitudes of earthquakes in both clusters were, on average, similar and relatively big. The same situation is in window 2 regarding waiting time and epicentral offset. The magnitudes of earthquakes in both clusters from that window were, on average, similar but small. In window 3, there are two clusters that repeat the pattern from window 2. However, the third, newly formed cluster grouped the earthquakes of short waiting times and epicentral offsets, and big magnitudes. In window 4, the pattern from window 2 is repeated.

The picture changes in window 5. Four clusters in this window grouped earthquakes of:

1. long waiting times and epicentral offsets, and big magnitudes,
2. long waiting times and epicentral offsets, and small magnitudes,
3. intermediate waiting times, short epicentral offsets, and big magnitudes,
4. short waiting times and epicentral offsets, and small magnitudes.

Five clusters from window 6 grouped earthquakes of:

1. long waiting times and epicentral offsets, and big magnitudes – the same characteristics as the one of the first cluster in window 5,
2. long waiting times and epicentral offsets, and small magnitudes – the same characteristics as the one of the second cluster in window 5,
3. long waiting times, short epicentral offsets, and small magnitudes,
4. short waiting times and epicentral offsets, and big magnitudes – the characteristics similar to that of the third cluster in window 5,
5. short waiting times and epicentral offsets, and small magnitudes – the same characteristics as the one of the fourth cluster in window 5.

Clustering of earthquakes in the last window 7 has the same form as that in window 6 except that the magnitudes of earthquakes in the third cluster of window 6 were small and in window 7 they were big.

This analysis indicates that earthquake clustering in the {*DT*, *DR*, *MC*} space was systematic. With the course of time, some characteristics of formed clusters continued while the other ones were steadily changing. The only radical change from August 2017 (window 4) to the main shock was the appearance of the additional cluster of earthquakes with long waiting times, short epicentral offset. On average, magnitudes of these earthquakes were at the beginning small and in the last window before the main shock became big.

## Sumamry and Conclusions

We studied clustering of earthquakes in the seismic series before the 2020 M7.0 Samos (Greece) main shock in the equivalent dimension space of the waiting time, epicentral offset, and magnitude. The study revealed that the average distance between these events was increasing steadily during the twelve–year period before the main shock. This premonitory signal was distinct and clearly significant.

These temporal changes in the average distance were driven by a systematic evolution of earthquake clustering in the used parameter space. Beginning from a two–cluster system, when the distance was minimal, the clustering development continued along two branches and ended before the main shock with the formation of five earthquake clusters of different characteristics.

Here we applied the same methodology of data analysis as the one used to study earthquake clustering before the 2017 M8.2 Tehuantepec (Mexico) main shock (Lasocki et al., 2024). The Tehuantepec earthquake, much stronger than the Samos main shock, exhibited a steep normal faulting (dip angle of 81°) that started at a depth of about 57 km and ruptured the entire lithosphere. The Samos earthquake type of faulting was also purely normal with typical dip angle (dip 37°) and focal depth of 12 km, which is again typical focal depth for strong ($M \geq 6.0$) crustal earthquakes in the Aegean where the seismogenic layer reaches up to 15 km. The tectonic setting of the Samos main shock and the seismic history of its site was much more complex than in the case of Tehuantepec earthquake.

Our present study show that despite the great differences between these two main shocks, both were preceded by distinct precursory time–changes of the average distance between earthquakes if these earthquakes were parameterized by the waiting time, epicentral offset, and magnitude. In both cases, there were no significant premonitory time–variations of earthquake clustering before the mainshocks when these variations were tracked in spaces of the original parameterization of earthquakes – the occurrence time, latitude, longitude, and magnitude.

This result has confirmed the capability of the use of the seismicity clustering process observed in the waiting time, epicentral offset, and magnitude parameter space for earthquake forecasting. It also suggests that the applicability of this approach may extend over wide range of main shock magnitudes and tectonic settings.

## Data and Resources

Data used in this work are taken from the catalogue of the Geophysics Department of the Aristotle University of Thessaloniki ( http://dx.doi.org/doi:10.7914/SN/HT) The completeness magnitude evaluation and the transformation to equivalent dimensions was done using the *Completeness Magnitude estimation* and *Transformation to Equivalent Dimensions* applications, implemented on the EPISODES platform – https://EpisodesPlatform.eu (Orlecka et al., 2020). Supplemental Material contains: Details on the transformation to equivalent dimensions applied in the present work; The selection of the criterion for dividing a set of points into optimal number of clusters; 3D distributions of earthquakes forming the clusters from the chains presented in Figure 5, whose centroid locations are shown in Figure 6. Geophysics Department contribution 000.

## Acknowledgements

This work was supported by a subsidy from the Polish Ministry of Education and Science for the Institute of Geophysics, Polish Academy of Sciences.

# References

Jordan, T.H., D. Schorlemmer, S. Wiemer, M. Gerstenberger, P. Maechling, M. Liukis, J. Zechar and the CSEP Collaboration (2007) Collaboratory for the Study of Earthquake Predictability (CSEP), Proc. of the 5th International Workshop of Statistical Seismology "Physical and Stochastic Modelling of Earthquake Occurrence and Forecasting." (StatSei 5), Erice, Italy, 31 May–6 June 2007

Karakostas VG, Papadimitriou EE, Karakaisis GF, Papazachos CB, Scordilis EM, Vargemezis G, Aidona E (2003) The 2001 Skyros, northern Aegean, Greece, Earthquake sequence: Off fault aftershocks, tectonic implications, and seismicity triggering. Geophys Res Lett 30(1): 1012, doi:10.1029/2002GL015814.

Karakostas VG, Papadimitriou EE, Tranos MD, Papazachos CB (2010) Active seismotectonic structures in the area of Chios Island, north Aegean Sea, revealed from microseismicity and fault plane solutions. Bull Geolog Soc Greece XLIII (3): 2064–2074.

Karakostas, V., Tan, O., Kostoglou, A., Papadimitriou, E. and Bonatis, P. (2021) Seismotectonic implications of the 2020 Samos, Greece, Mw7.0 mainshock based on high–relocation and source slip model. Acta Geophysica, https://doi.org/10.1007/s11600-021-00580-y

Kassaras I, Kapetanidis V, Ganas A, Tsanis A, Kosma C, Karakonstantis A, Valkaniotis S, Chailas S, Kouskouna V, Papadimitriou P (2020) The new seismotectonic atlas of Greece (v1.0) and its implementation. Geosciences 10: 447, https://doi:10.3390/geosciences10110447 .

Kato, A., and Y. Ben-Zion (2021) The generation of large earthquakes. Nature Reviews Earth & Environment 2(1), 26-39, https://doi.org/10.1038/s43017-020-00108-w

Lasocki, S. (2014). Transformation to equivalent dimensions—A new methodology to study earthquake clustering, Geophys J Int. 197, no. 2, 1224–1235, https://doi.org/10.1093/gji/ggu062

Lasocki, S., V. G. Karakostas, F. Ramón Zúñiga, E. E. Papadimitriou, and Y. Mahmood (2024). Premonitory Earthquakes Clustering Process in an Equivalent Dimensions Space before the 2017Mw 8.2 Tehuantepec, Mexico, Mainshock, Seismol. Res. Lett. XX, 1–13, https://doi.org/10.1785/0220240026

McClusky S, Balassanian S, Barka A, Demir C, Georgiev I, Hamburger M, Hurst K, Kahle H, Kastens K, Kekelidze G, King R, Kotzev V, Lenk O, Mahmoud S, Mishin A, Nadariya M, Ouzounis A, Paradisis D, Peter Y, Prilepi M, Reilinger R, Sanli I, Seeger H, Tealeb A, Toksoz MN, Veis G (2000) GPS constraints on crustal movements and deformations in the Eastern Mediterranean (1988–1997): Implications for plate dynamics. J Geophys Res 105: 5695–5719.

McKenzie DP (1972) Active tectonics of the Mediterranean region. Geophys J R Astron Soc 30: 109–185.

Mizrahi, L., Dallo, I., van der Elst, N. J., Christophersen, A., Spassiani, I., Werner, M. J., et al. (2024). Developing, testing, and communicating earthquake forecasts: Current practices and future directions. Reviews Geophysics, 62, e2023RG000823. https://doi.org/10.1029/2023RG000823

Papazachos BC, Comninakis PE (1971) Geophysical and tectonic features of the Aegean Arc. J. Geophys. Res. 76: 8517–8533.

Papazachos BC, Papazachou CC (2003) The earthquakes of Greece. Ziti Publication Co., Thessaloniki, pp. 304.

Papazachos BC, Papadimitriou EE, Kiratzi AA, Papazachos CB, Louvari EK (1998) Fault plane solutions in the Aegean Sea and the surrounding area and their tectonic implications. Boll Geof Teor Appl 39: 199–218.

Papazachos BC, Scordilis EM, Panagiotopoulos DG, Papazachos CB, Karakaisis GF (2004) Global relations between seismic fault parameters and moment magnitude of earthquakes. 10th Congr. Hellenic Geol. Soc., Thessaloniki, Greece, 14–17 April 2004, 539–540.

Tan O, Papadimitriou E, Pabuccu Z, Karakostas V, Yoruk A, Leptokaropoulos K (2014) A detailed analysis of microseismicity in Samos and Kusadasi (eastern Aegean Sea) areas (2014). Acta Geophys 62: 1283–1309, DOI: 10.2478/s11600–013–0194–1.

Wells DL, Coppersmith KJ (1994) New empirical relationships among magnitude, rupture length, rupture width, rupture area, and surface displacement. Bull Seism Soc Am 84: 974–1002.

Wessel P, Smith WHF, Scharroo R, Luis JF, Wobbe F (2013) Generic Mapping Tools: Improved version released. EOS Trans AGU 94: 409-410.

Zaliapin, I., and Y. Ben-Zion (2021). Perspectives on Clustering and Declustering of Earthquakes, Seismol. Res. Lett. 93, 386–401, https://doi.org/10.1785/0220210127

Authors' mailing addresses

Stanislaw Lasocki, Institute of Geophysics, Polish Academy of Sciences, 64 Księcia Janusza Str., 01-452 Warszawa, Poland lasocki@igf.edu.pl

Vasileios G. Karakostas, Geophysics Department, Aristotle University of Thessaloniki, GR54124 Thessaloniki, Greece, vkarak@geo.auth.gr

Eleftheria E. Papadimitriou, Geophysics Department, Aristotle University of Thessaloniki, GR54124 Thessaloniki, Greece, ritsa@geo.auth.gr